\documentclass[
 reprint,
 amsmath,amssymb,
 aps,
 pra,
 superscriptaddress,
]{revtex4-2}

\usepackage[section]{placeins}
\usepackage{graphicx}
\usepackage{dcolumn}
\usepackage{bm}
\usepackage{hyperref}
\hypersetup{hypertex=true,
colorlinks=true,
linkcolor=blue,
anchorcolor=blue,
citecolor=blue,
filecolor=blue,
urlcolor=blue,
citecolor=blue}
\usepackage[hyphenbreaks]{breakurl}

\begin{document}

\title{Multipole nuclear shielding factors of hydrogen atom confined by a spherical cavity}

\author{Yu Qing Dai}
\affiliation{College of Physics, Jilin University, Changchun 130012, People's Republic of China}

\author{Zhi Ling Zhou}
\affiliation{College of Physics, Jilin University, Changchun 130012, People's Republic of China}

\author{Henry E. Montgomery, Jr.}
\affiliation{Chemistry Program, Centre College, Danville, Kentucky 40422, USA}

\author{Yew Kam Ho}
\affiliation{Institute of Atomic and Molecular Sciences, Academia Sinica, Taipei 10617, Taiwan, Republic of China}

\author{Aihua Liu}\email{aihualiu@jlu.edu.cn}
\affiliation{Institute of Atomic and Molecular Physics, Jilin University, Changchun 130012, People's Republic of China}

\author{Li Guang Jiao}\email{lgjiao@jlu.edu.cn}
\affiliation{College of Physics, Jilin University, Changchun 130012, People's Republic of China}

\date{\today}


\begin{abstract}

Nuclear shielding factor is an important quantity to describe the response of an atom under the perturbation of an external field. In this work, we develop the sum-over-states numerical method and the Hylleraas variational perturbation approximation to calculate the multipole nuclear shielding factors for general one-electron systems and apply them to the model of the hydrogen atom confined by a spherical cavity. The generalized pseudospectral method is employed to solve the eigenstates of the unperturbed atom. The obtained dipole nuclear shielding factors are in good agreement with previous calculations and the higher-pole results are reported for the first time. The asymptotic behaviors of the multipole nuclear shielding factors in both the large- and small-confinement limits are analyzed with the assistance of variational perturbation theory. The free-atom values can be exactly reproduced by the second-order perturbation approximation and all multipole nuclear shielding factors in the small-confinement limit tend to zero by a linear law. The variational perturbation method manifests exponential convergence with increasing the order of approximation. The numerical and approximate methods developed in this work together pave the way for further investigation of the multipole nuclear shielding factors for general atomic systems.

\end{abstract}


\maketitle


\section{Introduction} \label{sec1}

The interaction of atoms with external electric fields or neighboring charged ions distorts the atomic charge distribution and induces electric multipole moments that are closely related to the atomic multipole polarizabilities \cite{Mitroy2010}. The distorted atomic charge, on the other hand, modifies the electric field inside the atom due to the screening effect of the atomic electrons. To describe such a complementary property between atoms and interacting external fields, Sternheimer \cite{SternheimerPR} first introduced a dimensionless quantity named the nuclear shielding factor, which for the dipole interaction is defined as the ratio of the change in the electric field at the nucleus due to the atomic charge distribution to the change in the electric field at the nucleus due to the external charge. From both the classical and quantum mechanical theories \cite{KaveeshwarJPB,FowlerMP}, it has been rigorously proved that for unconfined neutral atoms the electronic charge exactly cancels the external field at the nucleus, leading to a dipole nuclear shielding factor of unity. For an $N$-electron atomic system, the dipole nuclear shielding factor generalizes to $N/Z$ where $Z$ is the nuclear charge. Dalgarno \cite{DalgarnoAP} further proposed a unified definition of higher-pole nuclear shielding factors involving higher-order derivatives of the electric field. For instance, the quadrupole nuclear shielding factor is given by the ratio of the change in the gradient of the electric field at the nucleus caused by the atomic charge distribution to the change in the gradient of the electric field at the nucleus caused by the external charge. Compared to the multipole polarizabilities, where benchmark values for various atoms and ions have been established in the literature \cite{DalgarnoAP,Mitroy2010,Schwerdtfeger2019,Wang2021,Earwood2022}, multipole nuclear shielding factors have been less studied \cite{Cohen1966,Sternheimer1972,Sternheimer1974,Kolb1982,Devi1988,Komasa2001} and the reported values are far less accurate than those of polarizabilities.

In recent years, the investigation of confined quantum systems has attracted increasing interest due to their great importance in modeling, e.g., quantum dots in semiconductor materials, atoms and molecules trapped in zeolite or encaged in fullerenes, and charged ions immersed in strongly-coupled dense plasmas \cite{JaskolskiPR,BuchachenkoJPB,DolmatovRPC,SilAQC,JanevMRE,VasudevanJAC,LeyKoo2018,Aquilanti2021}. The simplest model of the spherically confined hydrogen atom, i.e., a hydrogen atom confined in an impenetrable spherical box, was originally proposed by Michels \textit{et al.} \cite{MichelsPhy} to study the spectrum of the hydrogen atom under high pressure and then was applied by Sommerfeld and Welker \cite{SommerfeldAP} to account for the delocalization of electronic wave functions in solids. Since then, this model atom has been extensively investigated by many authors focusing on a variety of structural and spectral properties, including the variation of energy spectrum and wave functions \cite{AguileraJPA,BanerjeeJCP,SenJCP,AquinoIJQC,RoyIJQC}, radial expectation values \cite{RoyIJQC,YuFBS,AquinoFBS,ZhouIJQC}, transition oscillator strengths \cite{StevanovicJPB,CabreraPRA,FloresJPB}, effective pressures \cite{Lude1977,Sen2002,Aquino2016,Zhou2023}, information-theoretic measures \cite{Sen2005,Patil2007,Aquino2013,Jiao2017}, etc. For the confined hydrogen atom interacting with external electric fields, the static and dynamic polarizabilities \cite{BanerjeeJCP,SenJCP,WaughJPB,MontgomeryPLA,HeEPJD,MukherjeePRA2021,MukherjeePRA2022} and hyperpolarizabilities \cite{BanerjeeJCP,WaughJPB,BhattacharyyaPLA,BhattacharyyaIJQC} have been thoroughly investigated and reported with high accuracy. The complementary quantities of nuclear shielding factors for confined atoms have also attracted considerable interest in the past years. Fowler \cite{FowlerMP} first investigated the dipole nuclear shielding factors for some confined one-electron systems, including the electron in a box, the hydrogen atom in a spherical box, and the harmonically bound electron in a box. Burrows and Cohen \cite{BurrowsPRA} formulated analytical closed-form solutions for the first-order perturbation corrections appropriate to the dipole interaction for the $s$-wave states of the confined hydrogen atom. The derived analytical expression of the dipole nuclear shielding factor nevertheless depends on numerical evaluations and integrations of hypergeometric Kummer functions. Laughlin \cite{LaughlinAQC} performed both numerical calculations and perturbation approximations of the ground state energies, wave functions, and dipole nuclear shielding factors for hydrogen-like atoms confined by spherical cavities. His numerical calculations are in good agreement with those of Burrows and Cohen \cite{BurrowsPRA} and the three perturbation approximations reproduce very well the asymptotic behavior of shielding factors in the large-, intermediate-, and small-box regions. It is also worth mentioning that Kaveeshwar \emph{et al.} \cite{KaveeshwarJPB} first defined the frequency-dependent dynamic dipole nuclear shielding factors for an atom subjected to a time-varying electric field. The recent work of Montgomery \emph{et al.} \cite{MontgomeryPS2017} undertook a detailed research on the dynamic dipole nuclear shielding factors for the plasma screened hydrogen atom confined in a spherical box. It has been shown that the boundary condition at the confinement radius plays a crucial role in modulating the nuclear shielding factors of confined atoms. A summary of previous work reveals that all of these studies focus only on the dipole nuclear shielding factors and they share the common theoretical foundations that the perturbed wave functions under external electric fields have to be explicitly calculated. The investigation of higher-pole terms for the confined hydrogen atom has not yet been reported so far.

In this work, we investigate the multipole nuclear shielding factors of the confined hydrogen atom using a sum-over-states formalism, which avoids the calculation of perturbed wave functions and works very well with standard numerical methods for solving the Schr\"{o}dinger equation. We further develop the Hylleraas variational perturbation theory (VPT) \cite{HylleraasZP,MontgomeryEPJH,MontgomeryPS2020,XiaPRE} to approximate the multipole nuclear shielding factors. Such a method provides a reasonably good and compact approximation that depends only upon the radial expectation values of the target state. Both methods are applied to calculate the multipole nuclear shielding factors of the confined hydrogen atom and aimed to provide accurate predictions for further reference. The following of this paper is organized as follows. In Sec. \ref{sec2}, we present in detail the calculation of multipole shielding factors in the sum-over-states formalism and the theoretical foundation in the VPT approximation. In Sec. \ref{sec3}, we present numerical calculations of multipole nuclear shielding factors for the confined hydrogen atom and compare the dipole results with previous predictions. The asymptotic behavior of shielding factors in both the large- and small-confinement limits and the convergence of the VPT method are also investigated. We summarize our work in Sec. \ref{sec4}. Unless otherwise stated, atomic units ($\hbar=m=e=1$) are used throughout this work.


\section{Theoretical method} \label{sec2}

\subsection{Multipole nuclear shielding factors}

The Schr\"{o}dinger equation for an unperturbed one-electron atom is given by
\begin{equation}\label{eq1}
H_{0} \left|\psi_{0i}\right\rangle \equiv \left[ -\frac{\nabla^{2}}{2} + V \right] \left|\psi_{0i}\right\rangle = E_{0i} \left|\psi_{0i}\right\rangle,
\end{equation}
where $E_{0i}$ and $\left|\psi_{0i}\right\rangle$ are the $i$th eigenenergy and wave function of the system, and $\left\{ \left| \psi_{0i} \right\rangle; i=1, 2, 3 ... \right\}$ forms a complete set of orthonormal basis functions
\begin{equation}
\label{eq2}
\left\langle\psi_{0i^{\prime}} \mid \psi_{0i}\right\rangle=\delta_{i^{\prime} i}.
\end{equation}
When the atom is confined in a spherical cavity, the potential term reads
\begin{equation}
\label{eq3}
V(r)=\left\{
\begin{aligned}
  &-\frac{Z}{r}  &\ \ (r < r_{\max}) \\
  &+\infty &\ \ (r \geq r_{\max})
\end{aligned}
\right. ,
\end{equation}
where $Z$ represents the nuclear charge of the atom and $r_{\max}$ the confinement radius, beyond which the wave function disappears, i.e., the system fulfills the Dirichlet boundary condition at $r_{\max}$. We keep in mind that the confined system reduces to the free hydrogenic ion in the limit of $r_{\max} \to \infty$.

\begin{figure}[!tbp]
\includegraphics[width=0.5\textwidth]{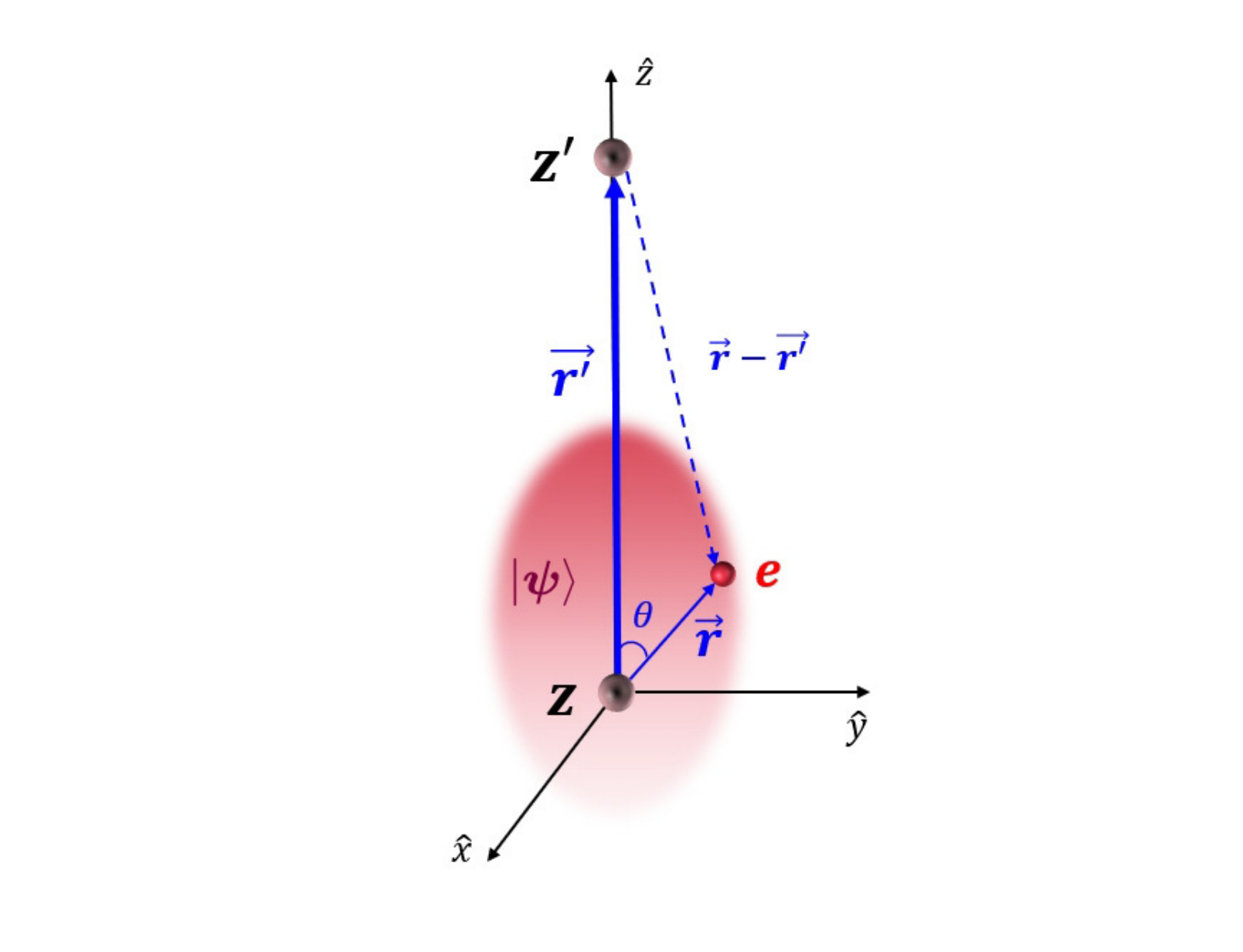}
\caption{\label{fig1} Schematic diagram of the coordinate system in investigating the nuclear shielding factors. An external charge $Z'$ is placed at the spherical coordinates $\mathbf{r'}$=($r'$, $0$, $0$), and the red area represents the distorted electronic wave function $\left|\psi \right\rangle$.}
\end{figure}

When an external charge $Z^{\prime}$ is placed at a far distance $r^{\prime}$ from the atomic nucleus ($r^{\prime}\gg r$, where $r$ is the atomic electron-nucleus distance), the electronic charge distribution of the atom changes due to the perturbation of the electric field generated by $Z^{\prime}$. If we choose the direction of $\mathbf{r'}$ as the $z$ axis of the coordinate (see Fig. \ref{fig1} for the schematic diagram of the system), the perturbation potential can be decomposed by applying a multipole expansion to the electron-$Z'$ interaction \cite{DalgarnoAP,LanghoffPR}
\begin{equation}
\label{eq4}
V(\mathbf{r},\mathbf{r'}) = -Z^{\prime} \sum_{k=1}^{\infty} \frac{r^k}{r^{\prime k+1}} P_k(\cos \theta) \equiv \sum_{k=1}^{\infty} \frac{Z'}{r^{\prime k+1}} v^{(k)},
\end{equation}
where $P_k(\cos \theta)$ is the Legendre polynomial of order $k$ and $\theta$ the polar angle of the electron. Here the $2^k$-pole $\mathbf{r}$-dependent potential is defined as $v^{(k)} \equiv -r^k P_k(\cos \theta)$. When the unperturbed system eigenstate $\left|\psi_{0i}\right\rangle$ is not degenerate, the exact wave function of the total Hamiltonian $H_0 + V(\mathbf{r},\mathbf{r'})$ can be expanded via the standard perturbation theory
\begin{equation}
\label{eq5}
\left|\psi_i \right\rangle = \left|\psi_{0i} \right\rangle +  \sum_{k=1}^{\infty} \frac{Z^{\prime}}{r^{\prime k+1}} \left|\psi _{1i} ^{(k)} \right\rangle + O(Z^{\prime 2}),
\end{equation}
where $\left|\psi _{1i} ^{(k)} \right\rangle$ is the first-order correction to the wave function of the $i$th eigenstate induced by the $2^k$-pole potential $v^{(k)}$, and it follows the first-order perturbed Schr\"{o}dinger equation
\begin{equation}
\label{eq6}
\left( H_0 -E_{0i} \right) \left|\psi_{1i}^{(k)}\right\rangle = \left( E_{1i}^{(k)} - v^{(k)} \right)  \left|\psi_{0i}\right\rangle,
\end{equation}
with
\begin{equation}
\label{eq7}
\left|\psi_{1i}^{(k)}\right\rangle = \sum\limits_{i^{\prime} \neq i} \frac{ \left\langle \psi_{0i^{\prime}} \left| v^{(k)} \right| \psi_{0i} \right\rangle }{E_{0i}-E_{0i^{\prime}}}\left|\psi_{0i^{\prime}}\right\rangle,
\end{equation}
and
\begin{equation}
\label{eq8}
E_{1i}^{(k)} = \left\langle \psi_{0i} \left| v^{(k)} \right| \psi_{0i} \right\rangle.
\end{equation}

The combined electrostatic potential at the nucleus due to the atomic electron and to the external charge $Z'$ is
\begin{equation}
\label{eq9}
\mathcal{V}(\mathbf{r},\mathbf{r'}) = V(r) + V (r^{\prime}) \equiv -\frac{Z}{r} + \frac{Z Z^{\prime}}{r^{\prime}}.
\end{equation}
The electric field along the $z$ axis corresponding to the above electrostatic potential, averaged over the electronic wave function $\left|\psi_i \right\rangle$, is therefore given by
\begin{equation}
\label{eq10}
\mathcal{E}(z,z') = \left\langle \psi_{i} \left| \frac{P_1(\cos \theta)}{r^{2}} \right| \psi_{i} \right\rangle - \frac{Z^{\prime}}{r^{\prime 2}}.
\end{equation}
Substituting Eq. (\ref{eq5}) into Eq. (\ref{eq10}) and retaining only the first-order term in $Z'$, one gets
\begin{equation}
\label{eq11}
\mathcal{E}(z,z') \approx 2 \frac{Z^{\prime}}{r^{\prime 2}} \left\langle \psi_{1i}^{(1)} \left| \frac{P_1(\cos \theta)}{r^{2}} \right| \psi_{0i} \right\rangle - \frac{Z^{\prime}}{r^{\prime 2}}.
\end{equation}
The dipole nuclear shielding factor is then defined as the ratio between the magnitudes of the first and second terms on the right hand side of Eq. (\ref{eq11})
\begin{equation}
\label{eq12}
\gamma^{(1)} = 2 \left\langle \psi_{1i}^{(1)} \left| \frac{P_1(\cos \theta)}{r^{2}} \right| \psi_{0i} \right\rangle.
\end{equation}
In the literature, the dipole nuclear shielding factor was usually referred to as $\beta_\infty$. Here we employ the notation $\gamma$ for the convenience of defining multipole terms \cite{DalgarnoAP}.

The $2^{k}$-pole nuclear shielding factor can be defined in a similar way by performing the $k$th-order derivative of the electric fields induced by the atomic electron and by the external charge, respectively, and averaged over the atomic charge distribution
\begin{equation}
\label{eq13}
\gamma^{(k)}= - \frac{ \left\langle\psi_i\left| \partial^k V(r) / \partial z^k\right| \psi_i\right\rangle }{\partial^k V\left(r^{\prime}\right) / \partial z^{\prime k}},
\end{equation}
where the minus sign on the right hand side is simply due to the fact that the nuclear shielding factors in this situation are positively defined. Substituting Eq. (\ref{eq5}) into Eq. (\ref{eq13}) and making the first-order approximation leads to
\begin{equation}
\label{eq14}
\gamma^{(k)}= - 2\frac{Z^{\prime}}{r^{\prime k+1}} \frac{ \left\langle \psi_{1i}^{(k)} \left| \partial^k V(r) / \partial z^k \right| \psi_{0i} \right\rangle }{\partial^k V\left(r^{\prime}\right) / \partial z^{\prime k}}.
\end{equation}
After simple manipulations with the partial derivatives of the corresponding potentials
\begin{equation}
\label{eq15}
\frac{\partial^k V\left(r\right)}{\partial z^k}=(-1)^{k+1}  k! \frac{P_k(\cos \theta)}{r^{ k+1}},
\end{equation}
\begin{equation}
\label{eq16}
\frac{\partial^k V\left(r^{\prime}\right)}{\partial z^{\prime k}}=(-1)^k  k! \frac{Z^{\prime}}{r^{\prime k+1}},
\end{equation}
one arrives at the $2^{k}$-pole nuclear shielding factor for a general one-electron atom with Coulomb interaction
\begin{equation}
\label{eq17}
\gamma^{(k)}=2\left\langle \psi_{1i}^{(k)}\left|\frac{P_k(\cos \theta)}{r^{ k+1}}\right| \psi_{0i} \right\rangle .
\end{equation}

The first-order correction to the system wave function was given in Eq. (\ref{eq7}) and the substitution of it into the above equation finally yields the $2^{k}$-pole nuclear shielding factor in the sum-over-states formalism
\begin{widetext}
\begin{equation}
\label{eq18}
\gamma^{(k)}=2 \sum_{i^{\prime}\neq i} \frac{\left\langle\psi_{0i}|r^{k} P_k(\cos \theta)| \psi_{0i^{\prime}}\right\rangle\left\langle\psi_{0i^{\prime}}\left|r^{-(k+1)}P_k(\cos \theta)\right| \psi_{0i}\right\rangle}{E_{0i^{\prime}}-E_{0i}}.
\end{equation}
\end{widetext}
If the initial and intermediate unperturbed states are labeled by $\left|nlm\right\rangle$ and $\left|n^{\prime} l^{\prime} m^{\prime}\right\rangle$, respectively, the transition matrix elements in the numerator can be further decomposed into radial and angular parts
\begin{widetext}
\begin{equation}
\label{eq19}
\left\langle n^{\prime} l^{\prime} m^{\prime}|r^{q} P_k(\cos \theta)| n l m\right\rangle\\
= (-1)^{m^{\prime}} \langle r^{q}\rangle_{n l}^{n^{\prime} l^{\prime}} \sqrt{(2 l+1)\left(2 l^{\prime}+1\right)}
\left(\begin{array}{ccc}
l^{\prime} & k & l \\
0 & 0 & 0
\end{array}\right)
\left(\begin{array}{ccc}
l^{\prime} & k & l \\
-m' & 0 & m
\end{array}\right),
\end{equation}
\end{widetext}
where the radial part is given by
\begin{equation}
\label{eq20}
\langle r^{q}\rangle_{n l}^{n^{\prime} l^{\prime}}=\int_0^{r_{\max }} R_{n^{\prime} l^{\prime}}(r) r^{q} R_{n l}(r) r^2 dr.
\end{equation}
Here $R_{n^{\prime} l^{\prime}}(r)$ and $R_{n l}(r)$ are the corresponding radial components of wave functions for the $\left|nlm\right\rangle$ and $\left|n^{\prime} l^{\prime} m^{\prime}\right\rangle$ states, respectively.


\subsection{Generalized pseudospectral method}

In this work, we employ the generalized pseudospectral (GPS) method to solve the unperturbed Schr\"{o}dinger equation of Eq. (\ref{eq1}). Details about the GPS method are available elsewhere \cite{YaoCPL,ChuPR,CanutoSM,ShenSM,DeloffAP,SenPLA,ZhuIJQC} and here we only give a brief summary of it. The radial variable $r \in\left[0, r_{\max }\right]$ is first mapped onto a finite interval of $x \in\left[-1, 1\right]$ by utilizing a rational mapping function
\begin{equation}
\label{eq21}
r \equiv f(x)=L\frac{1+x}{1-x+\alpha },
\end{equation}
where $\alpha = 2L/r_{\max}$ and $L$ is an adjustable mapping parameter. The radial wave function is then transformed into a function about the new variable $x$ via
\begin{equation}
\label{eq22}
\phi_{nl} (x) = \sqrt{f^{\prime}(x)} r R_{nl}(r).
\end{equation}

The variable $x$ is discretized based on the Legendre-Gauss-Lobatto quadrature with abscissas (collocation points) $x_{j(j=0, \dots, N)}$ and weights $\omega_j$ \cite{ShenSM}. The spectral approximation expresses the unknown function $\phi (x)$ by the values of the function itself at all collocation points
\begin{equation}
\label{eq23}
\phi(x) \approx \sum_{j=0}^{N} g_{j}(x) \phi(x_{j}),
\end{equation}
where the cardinal function $g_j(x)$ reproduces the Dirac delta function at collocation points, i.e., $g_j(x_i)=\delta_{ij}$. The substitution of the spectral approximation of the wave function into the radial Schr\"{o}dinger equation leads to a standard eigenvalue problem. After solving it, one obtains the system energy spectrum including both bound and pseudo-continuum states and corresponding wave functions in the discrete variable representation. As a result, the integration in Eq. (\ref{eq20}) can be conveniently calculated via a summation over the collocation points
\begin{equation}
\label{eq24}
\langle r^{q}\rangle_{n l}^{n^{\prime} l^{\prime}}=\sum_{j=0}^N f^{q}(x_j) \phi_{n^{\prime} l^{\prime}}(x_j) \phi_{n l}(x_j) \omega_j.
\end{equation}
It has been well-established in the literature \cite{YaoCPL,HePRE} that the GPS method shows exponentially fast convergence with respect to increasing the dimension of discretization $N$ and holds extremely high numerical stability at large values of $N$. It has also been demonstrated that the method depends weakly on the mapping parameter $L$ provided $N$ is large \cite{JiaoJPB}. Furthermore, the obtained eigenstates form a near complete set of orthonormal basis functions that are suitable for the sum-over-states calculations \cite{ZhuPP,HePRE}.


\subsection{Variational perturbation theory}

Besides accurate numerical calculations based on the sum-over-states method, the Hylleraas VPT method \cite{HylleraasZP,MontgomeryEPJH,MontgomeryPS2020,XiaPRE} provides a compact and easy-to-use analytical formula to approximate the nuclear shielding factors. Analogous work has been done on the dipole polarizability of one-electron atoms, leading to the well-known Kirkwood (first-order) \cite{Kirkwood} and Buckingham (second-order) \cite{Buckingham} approximations. In the present work, we develop the VPT approximation for the generalized $2^{k}$-pole nuclear shielding factors of one-electron atoms.

We focus on the ground state of the system, where the wave functions $\left|\psi_{0i}\right\rangle$ and $\left|\psi_{0i^{\prime}}\right\rangle$ in Eq. (\ref{eq18}) are simplified as $\left|\psi_{0}\right\rangle$ and $\left|\psi_{m}\right\rangle$, respectively. We keep in mind that both of them are eigenstates of the unperturbed atomic Hamiltonian. As a result, the sum-over-states formalism of $\gamma^{(k)}$ in Eq. (\ref{eq18}) can be rewritten as
\begin{equation}
\label{eq25}
\gamma^{(k)} = 2 \sum_{m\neq 0} \frac{\left\langle\psi_0|r^{k} P_k(\cos \theta)| \psi_m\right\rangle\left\langle\psi_m\left|r^{-(k+1)}P_k(\cos \theta)\right| \psi_0\right\rangle}{E_m-E_0}.
\end{equation}
We further define a projection operator on the subspace of functions that are orthogonal to the ground state
\begin{equation}
\label{eq26}
\hat{P}_{0} = 1 - \left|\psi_0\right\rangle\left\langle\psi_0\right| = \sum_{m \neq 0}\left| \psi_m\right\rangle\left\langle\psi_m\right|,
\end{equation}
then Eq. (\ref{eq25}) can be formally written as
\begin{equation}
\label{eq27}
\gamma^{(k)} = 2 \left\langle\psi_0\left|r^{k} P_k(\cos \theta) \hat{P}_{0} \left(H - E_0\right)^{-1} \hat{P}_{0} r^{-(k+1)}P_k(\cos \theta) \right| \psi_0\right\rangle.
\end{equation}

The key step in applying the Hylleraas VPT method is to re-expand the subspace expanded by $\left\{ \left|\psi_{m}\right\rangle; m \ne 0 \right\}$ using a reduced $J$-dimensional basis set
\begin{equation}
\label{eq28}
\eta_p^{(k)} = \hat{P}_{0} \xi_p^{(k)} \ \  (p=1, 2, \dots J),
\end{equation}
where the functions $\xi_p^{(k)}$ have the same symmetry as $r^{k} P_k(\cos \theta)\psi_0$.
For simplicity, $\xi_p^{(k)}$ can be chosen as
\begin{equation}
\label{eq29}
\xi_p^{(k)} = r^{k} P_k(\cos \theta) r^{p-1}\psi_0.
\end{equation}
Substituting Eq. (\ref{eq29}) into (\ref{eq28}) and applying the spectral decomposition of the Hamiltonian operator in Eq. (\ref{eq27}) based on the reduced $J$-dimensional basis set $\eta^{(k)}$, one obtains the matrix form of the $J$th-order approximation of the $2^{k}$-pole nuclear shielding factors
\begin{widetext}
\begin{equation}
\label{eq30}
\gamma_{J}^{(k)} = 2[\mathbf{B}^{(k)}]^{\dagger}[\mathbf{A}^{(k)}]^{-1}[\mathbf{C}^{(k)}]
  = 2\left(\begin{array}{llll}
  B_1^{(k)} & B_2^{(k)} & \cdots & B_J^{(k)}
  \end{array}\right)\left(\begin{array}{cccc}
  A_{11}^{(k)} & A_{12}^{(k)} & \cdots & A_{1 J}^{(k)} \\
  A_{21}^{(k)} & A_{22}^{(k)} & \cdots & A_{2 J}^{(k)} \\
  \vdots & \vdots & \ddots & \vdots \\
  A_{J 1}^{(k)} & A_{J 2}^{(k)} & \cdots & A_{J J}^{(k)}
  \end{array}\right)^{-1}\left(\begin{array}{c}
  C_1^{(k)} \\
  C_2^{(k)} \\
  \vdots \\
  C_J^{(k)}
  \end{array}\right),
\end{equation}
\end{widetext}
where $[\mathbf{A}^{(k)}]$ is a $J \times J$ matrix with its elements given by
\begin{widetext}
\begin{equation}
\label{eq31}
A_{pq}^{(k)} = \left\langle \eta_p^{(k)} \left| H-E_{0} \right| \eta_q^{(k)}\right\rangle
= \frac{(k+p-1)(k+q-1)+k(k+1)}{2(2k+1)}\left\langle r^{2k+p+q-4} \right\rangle,
\end{equation}
\end{widetext}
and $[\mathbf{B}^{(k)}]$ and $[\mathbf{C}^{(k)}]$ are both $J$-dimensional column vectors with matrix elements
\begin{equation}
\label{eq32}
B_p^{(k)} = \left\langle \eta_p^{(k)} \left| r^{k}P_k(\cos \theta)      \right| \psi_0\right\rangle = \frac{1}{2k+1}\left\langle r^{2k+p-1} \right\rangle,
\end{equation}
\begin{equation}
\label{eq33}
C_p^{(k)} = \left\langle \eta_p^{(k)} \left| r^{-(k+1)}P_k(\cos \theta) \right| \psi_0\right\rangle = \frac{1}{2k+1}\left\langle r^{p-2}    \right\rangle.
\end{equation}

For $J=1$, Eq. (\ref{eq30}) reduces to the first-order approximation of the $2^{k}$-pole nuclear shielding factors
\begin{equation}
\label{eq34}
\gamma^{(k)}_1 = 2 \frac{B_1C_1}{A_{11}} = \frac{4}{k(2k+1)^2} \frac{\left\langle r^{-1} \right\rangle \left\langle r^{2k} \right\rangle}{\left\langle r^{2k-2} \right\rangle},
\end{equation}
and for $J=2$ to the second-order approximation of $\gamma^{(k)}$ in a more complex form
\begin{widetext}
\begin{equation}
\begin{split}
\label{eq35}
&\gamma^{(k)}_2  = 2 \frac{B_2A_{11}C_2 -B_1A_{12}C_2 - B_2A_{21}C_1 + B_1A_{22}C_1}{A_{11}A_{22}-A_{12}A_{21}} \\
& = \frac{ 4 \left[ k(2k+1) \left \langle r^{2k-2} \right \rangle \left \langle r^{2k+1} \right \rangle - 2k(k+1) \left \langle r^{2k-1} \right \rangle \left \langle r^{2k} \right \rangle
- 2k(k+1) \left \langle r^{-1} \right \rangle \left \langle r^{2k-1} \right \rangle \left \langle r^{2k+1} \right \rangle + (k+1)(2k+1) \left \langle r^{-1} \right \rangle \left \langle r^{2k} \right \rangle^2 \right] }{ k(k+1)(2k+1) \left[ (2k+1)^2 \left \langle r^{2k-2} \right \rangle \left \langle r^{2k} \right \rangle - 4k(k+1) \left \langle r^{2k-1} \right \rangle^2 \right] }.
\end{split}
\end{equation}
\end{widetext}
Taking the dipole nuclear shielding factor as an example, the substitution of $k=1$ into Eqs. (\ref{eq34}) and (\ref{eq35}) results in, respectively, the first-order approximation of
\begin{equation}
\label{eq36}
\gamma^{(1)}_1 = \frac{4}{9} \left\langle r^{-1} \right\rangle \left\langle r^2 \right\rangle,
\end{equation}
and the second-order approximation of
\begin{widetext}
\begin{equation}
\begin{split}
\label{eq37}
\gamma^{(1)}_2  =  \frac{2 \left[ 3 \left \langle r^3 \right \rangle - 4 \left \langle r \right \rangle \left \langle r^2 \right \rangle - 4 \left \langle r^{-1} \right \rangle \left \langle r \right \rangle \left \langle r^3 \right \rangle + 6 \left \langle r^{-1} \right \rangle \left \langle r^2 \right \rangle^2 \right] }{3 \left[ 9 \left \langle r^2 \right \rangle - 8 \left \langle r \right \rangle^2 \right] }.
\end{split}
\end{equation}
\end{widetext}
To a certain extent, they are analogous to, respectively, the Kirkwood \cite{Kirkwood} and Buckingham \cite{Buckingham} approximations of the dipole polarizabilities for the ground state of one-electron atoms. We noticed that both $\gamma^{(1)}_1$ and $\gamma^{(1)}_2$ have been derived by Fowler \cite{FowlerMP}, but differ from the present forms by a minus sign. On the contrary, Eqs. (\ref{eq36}) and (\ref{eq37}) are always positively defined. We finally summarize that the combination of Eqs. (\ref{eq30}) to (\ref{eq33}) derived in the present work provides a unified VPT approximation of the generalized $2^{k}$-pole nuclear shielding factors to arbitrary orders. Furthermore, as we will show later, this approximation serves as a useful tool to analyze the asymptotic behavior of nuclear shielding factors in both the large- and small-confinement limits.


\section{Results and discussion} \label{sec3}

The $2^k$-pole nuclear shielding factors for the ground state of the confined hydrogen atom at some selected values of $r_{\max}$ are presented in Table \ref{tab1}, where $k=1$, $2$, $3$, and $4$ represent the dipole, quadrupole, octupole, and hexadecapole nuclear shielding factors, respectively. The sum-over-states method in Eq. (\ref{eq18}) was employed and the unperturbed Hamiltonian was solved by utilizing the GPS method with $N=100$ and $L=100$ and $10 r_{\max}$ for the free and confined situations, respectively. All results shown in Table \ref{tab1} are converged to the last reported digit. This can be seen from the free hydrogen atom with $r_{\max} = \infty$, where Dalgarno \cite{DalgarnoAP} has rigorously derived
\begin{equation}
\label{eq38}
\gamma^{(k)} (r_{\max} =\infty) = \frac{2}{k(k+1)}.
\end{equation}
It is interestingly found that for the free hydrogen atom the second-order VPT approximation given in Eq. (\ref{eq35}) exactly reproduces the above analytical result. Recalling the ground state wave function of the hydrogen atom
\begin{equation}
\label{eq39}
\psi_0 = \frac{1}{\sqrt{\pi } } e^{-r},
\end{equation}
and corresponding radial expectation values
\begin{equation}
\label{eq40}
\left\langle r^p \right\rangle = \frac{(p+2)!}{2^{(p+1)}}; \ \ (p \geq -2),
\end{equation}
it can be readily obtained from Eqs. (\ref{eq34}) and (\ref{eq35}) that
\begin{equation}
\label{eq41}
\gamma^{(k)}_1 = \frac{2(k+1)}{k(2k+1)},
\end{equation}
and
\begin{equation}
\label{eq42}
\gamma^{(k)}_2 = \frac{2}{k(k+1)}.
\end{equation}
This is similar to the situation of the polarizabilities for the free hydrogen atom, where the second-order VPT approximation reproduces exactly the analytical solution of multipole polarizabilities \cite{XiaPRE}. Another situation that exhibits analytical solution of nuclear shielding factors is the confined hydrogen atom at $r_{\max} = 2$ where the so-called \textit{incidental degeneracy} appears between the ground state of the confined hydrogen atom and the $2s$ state of the free hydrogen atom, i.e., $E_{1s}(r_{\max} =2) = E_{2s}(r_{\max} =\infty) = -0.125$ \cite{SenJCP}. This is caused by an incidental coincidence of the confinement radius with one of nodes of the eigenstates of the free hydrogen atom. The dipole nuclear shielding factor for this special case has been derived by Laughlin \cite{LaughlinAQC}
\begin{equation}
\label{eq43}
\gamma^{(1)} (r_{\max} =2) = \frac{12 - 88 e^{-2}}{3(1 - 7 e^{-2})}.
\end{equation}
The present numerical calculation via the sum-over-states method reproduces this value almost near the working precision. Because incidental degeneracy is a rather common phenomenon in spherically confined systems, it is possible that analytical solutions of multipole nuclear shielding factors probably exist among a wide variety of incidentally degenerate states.

Table \ref{tab1} also includes the comparison of the present results with previous calculations from Burrows and Cohen \cite{BurrowsPRA}, Montgomery and Pupyshev \cite{MontgomeryPS2017}, and Laughlin \cite{LaughlinAQC}, for the dipole nuclear shielding factor only. Burrows and Cohen \cite{BurrowsPRA} calculated the dipole nuclear shielding factor for the ground state of the confined hydrogen atom utilizing two equivalent expressions, where both the ground state wave function and its first-order perturbed wave function were obtained in analytical forms. Their two calculations are almost identical to eight significant decimal digits. Montgomery and Pupyshev \cite{MontgomeryPS2017} solved the first-order correction of wave function using a power-series decomposition numerical method. The obtained results are in good agreement with those of Burrows and Cohen \cite{BurrowsPRA} within the reported significant digits. Laughlin \cite{LaughlinAQC} investigated the dipole nuclear shielding factor for the confined hydrogen atom with both numerical method and three approximations, i.e., the asymptotic expansion at large-box radii and the perturbation theories at intermediate- and small-box radii. The three approximations proposed by Laughlin \cite{LaughlinAQC} show good agreement with numerical results within the corresponding range of application, while discrepancy increases when the confinement radius is far beyond the effective range. Taking into account possible rounding errors, the present dipole results are in complete agreement with all previous numerical calculations mentioned above. To the best of our knowledge, no comparison can be made on the higher-order nuclear shielding factors.

\begin{table*}[!htbp]
\caption{\label{tab1} The $2^k$-pole ($k=1, 2, 3, 4$) nuclear shielding factors for the ground state of spherically confined hydrogen atoms at some selected values of $r_{\max}$. Numbers in parentheses represent powers of ten.}
\begin{ruledtabular}
\setlength\tabcolsep{3pt}
\begin{tabular}{lllll}
\multicolumn{1}{c}{$r_{\max}$} & \multicolumn{1}{c}{$\gamma^{(1)}$ } & \multicolumn{1}{c}{$\gamma^{(2)}$ } & \multicolumn{1}{c}{$\gamma^{(3)}$ } & \multicolumn{1}{c}{$\gamma^{(4)}$ } \\
\hline
$\infty $ & 1.0000000000000000000    & 3.3333333333333333333(-1)& 1.6666666666666666666(-1)& 1.0000000000000000000(-1)\\
20        & 9.9999999999838180004(-1)& 3.3333333333150407526(-1)& 1.6666666666430344569(-1)& 9.9999999996867779968(-2)\\
18        & 9.9999999994286330478(-1)& 3.3333333327383773089(-1)& 1.6666666659582704029(-1)& 9.9999999913196225774(-2)\\
          & 1.000000000\footnotemark[3],1.000000000\footnotemark[4] \\
14        & 9.9999994032361584007(-1)& 3.3333328183907850946(-1)& 1.6666661558318981433(-1)& 9.9999947334671050125(-2)\\
          & 9.99999940(-1)\footnotemark[3],9.99999940(-1)\footnotemark[4] \\
12        & 9.9999830179309484715(-1)& 3.3333201948683413764(-1)& 1.6666548999207512715(-1)& 9.9998895946226175181(-2)\\
          & 9.99998302(-1)\footnotemark[3],9.99998303(-1)\footnotemark[4] \\
10        & 9.9995758584767508255(-1)& 3.3330429920623135174(-1)& 1.6664337223699546439(-1)& 9.9980186741455312160(-2)\\
          & 9.99957586(-1)\footnotemark[1],9.9996(-1)\footnotemark[2] \\
          & 9.99957586(-1)\footnotemark[3],9.99957628(-1)\footnotemark[4] \\
8         & 9.9913157325707052943(-1)& 3.3281579157744676164(-1)& 1.6629710569046865880(-1)& 9.9715266649857197975(-2)\\
          & 9.99131573(-1)\footnotemark[1],9.9913(-1)\footnotemark[2] \\
          & 9.99131573(-1)\footnotemark[3],9.991339(-1)\footnotemark[4] \\
7         & 9.9648943256217795703(-1)& 3.3139435441966356043(-1)& 1.6536187829906699916(-1)& 9.9041550841267739050(-2)\\
6         & 9.8727417009910006043(-1)& 3.2684573622751671734(-1)& 1.6254881262901082261(-1)& 9.7107849233401805898(-2)\\
          & 9.87274170(-1)\footnotemark[1],9.8727(-1)\footnotemark[2] \\
          & 9.87274170(-1)\footnotemark[3],9.8756(-1)\footnotemark[4] \\
          & 1.11621475\footnotemark[5],1.019625\footnotemark[6] \\
5         & 9.6001714211924413850(-1)& 3.1456769190210761966(-1)& 1.5539908752697261918(-1)& 9.2399775256802924304(-2)\\
          & 9.60017142(-1)\footnotemark[3],9.65866521(-1)\footnotemark[5] \\
          & 9.60170(-1)\footnotemark[6]  \\
4         & 8.9447312051058546173(-1)& 2.8771325993838978114(-1)& 1.4061966318090481191(-1)& 8.3028995288463884053(-2)\\
          & 8.94473121(-1)\footnotemark[1],8.9447(-1)\footnotemark[2] \\
          & 8.94473121(-1)\footnotemark[3],9.27(-1)\footnotemark[4] \\
          & 8.94592798(-1)\footnotemark[5],8.94475(-1)\footnotemark[6]\\
3         & 7.6880420176134238075(-1)& 2.4086157735590673643(-1)& 1.1611263508196333182(-1)& 6.7981085083176809819(-2)\\
          & 7.68804022(-1)\footnotemark[3],7.68804934(-1)\footnotemark[5] \\
          & 7.68804(-1)\footnotemark[6]\\
2         & 5.7290211617784767632(-1)& 1.7405126543689635397(-1)& 8.2668579418240656159(-2)& 4.7979705978422700487(-2)\\
          & 5.72902116(-1)\footnotemark[1],5.7290(-1)\footnotemark[2] \\
          & 5.72902116(-1)\footnotemark[3],5.72902117(-1)\footnotemark[5] \\
          & 5.72902(-1)\footnotemark[6]\\
1         & 3.1221176945913015396(-1)& 9.1934682925915109154(-2)& 4.3045725203590657008(-2)& 2.4780235814157901180(-2)\\
          & 3.12211769(-1)\footnotemark[1],3.1221(-1)\footnotemark[2] \\
          & 3.12211769(-1)\footnotemark[3],3.12211769(-1)\footnotemark[5] \\
0.5       & 1.6166057660613689205(-1)& 4.6896858485406278956(-2)& 2.1812005883532792304(-2)& 1.2509622183206869106(-2)\\
          & 1.61660577(-1)\footnotemark[1]\\
          & 1.61660577(-1)\footnotemark[3],1.61660577(-1)\footnotemark[5] \\
0.25      & 8.2114602177907379751(-2)& 2.3649698612624173227(-2)& 1.0964489620263470532(-2)& 6.2771397613556343678(-3)\\
          & 8.21146022(-2)\footnotemark[1]\\
          & 8.2114602(-2)\footnotemark[3],8.2114602(-2)\footnotemark[5]\\
0.125     & 4.1365980962590358945(-2)& 1.1871668864173085081(-2)& 5.4953514675319982381(-3)& 3.1433353431026201414(-3)\\
          & 4.13659810(-2)\footnotemark[1]\\
          & 4.1365981(-2)\footnotemark[3],4.1365981(-2)\footnotemark[5]\\
0.1       & 3.3141400769439843166(-2)& 9.5046134610282869288(-3)& 4.3982870768787619475(-3)& 2.5153819342441184175(-3)\\
0.07      & 2.3239582912776822128(-2)& 6.6592792297077068375(-3)& 3.0804586121756046950(-3)& 1.7613522542646073231(-3)\\
0.05      & 1.6618935321924071390(-2)& 4.7594814346222884638(-3)& 2.2011052611403661016(-3)& 1.2583810495411879348(-3)\\
0.01      & 3.3314320850661939256(-3)& 9.5302463147208564711(-4)& 4.4052659696905096413(-4)& 2.5178218581641488981(-4)\\
0.007     & 2.3324020149701554586(-3)& 6.6717599423904806092(-4)& 3.0838445305780027372(-4)& 1.7625298430144247443(-4)\\
0.005     & 1.6661916039579511736(-3)& 4.7658223209652621727(-4)& 2.2028213856767397055(-4)& 1.2589758002338242019(-4)\\
0.001     & 3.3331433879922185594(-4)& 9.5327613795786172445(-5)& 4.4059433983478020839(-5)& 2.5180549616342419953(-5)\\
0.0007    & 2.3332402630458022470(-4)& 6.6729915462272501720(-5)& 3.0841761413644855550(-5)& 1.7626438887699587034(-5)\\
0.0005    & 1.6666191828225210480(-4)& 4.7664504258386955202(-5)& 2.2029904629513310272(-5)& 1.2590339273914328482(-5)\\
0.0001    & 3.3333143405927854359(-5)& 9.5330124101516431890(-6)& 4.4060109398217550891(-6)& 2.5180781649859278007(-6)\\
\end{tabular}
\end{ruledtabular}
\footnotetext[1]{Burrows and Cohen \cite{BurrowsPRA}.}
\footnotetext[2]{Montgomery and Pupyshev \cite{MontgomeryPS2017}.}
\footnotetext[3]{Laughlin (numerical calculations) \cite{LaughlinAQC}.}
\footnotetext[4]{Laughlin (large-box first-order expansion) \cite{LaughlinAQC}.}
\footnotetext[5]{Laughlin (small-box perturbation theory) \cite{LaughlinAQC}.}
\footnotetext[6]{Laughlin (intermediate-box perturbation theory) \cite{LaughlinAQC}.}
\end{table*}

\begin{figure}[!tbp]
\includegraphics[width=0.5\textwidth]{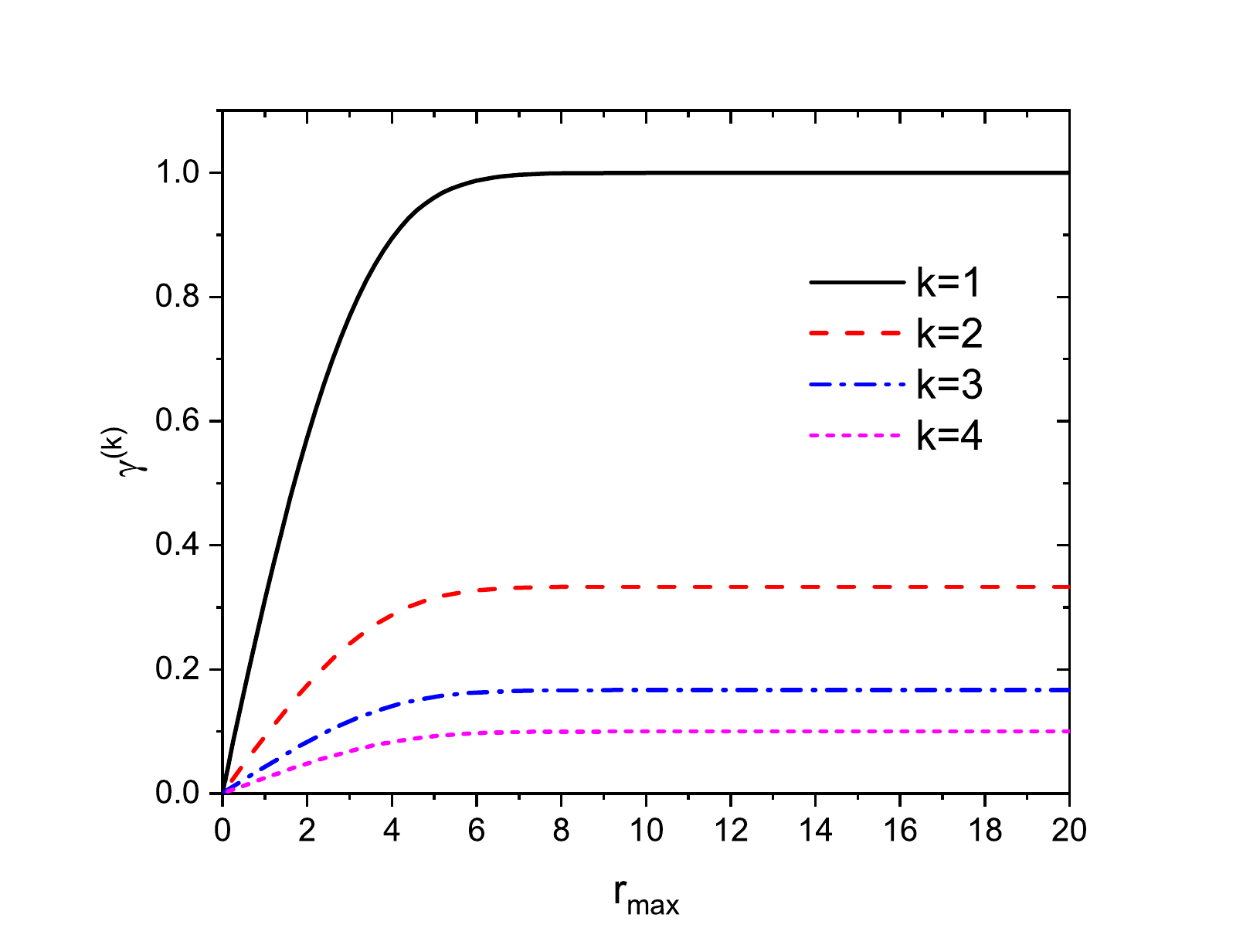}
\caption{\label{fig2} Variation of the $2^k$-pole ($k=1, 2, 3, 4$) nuclear shielding factors with the confinement radius $r_{\max}$ for the ground state of the spherically confined hydrogen atom.}
\end{figure}

For a better understanding of the variation of multipole nuclear shielding factors with respect to the confinement radius, we depict in Fig. \ref{fig2} the dipole, quadrupole, octupole, and hexadecapole results for the ground state of the confined hydrogen atom with $r_{\max}$ decreasing from $20$ to zero. It can be seen from the figure that the nuclear shielding factors approach the free-atom exact solution of $2/[k(k+1)]$ at large values of $r_{\max}$. They gradually decrease as the confinement radius becomes smaller than $5$ and eventually tend to zero in the limit of $r_{\max} \to 0$. Such a trend coincides with the commonsense knowledge that as the spatial confinement compresses the atom into a smaller region near the nucleus, the atomic electrons produce less shielding effect on the external charge. In this sense, the nuclear shielding factors provide complementary information on atoms influenced by and at the same time modifying the external field.

\begin{figure}[!tbp]
\includegraphics[width=0.5\textwidth]{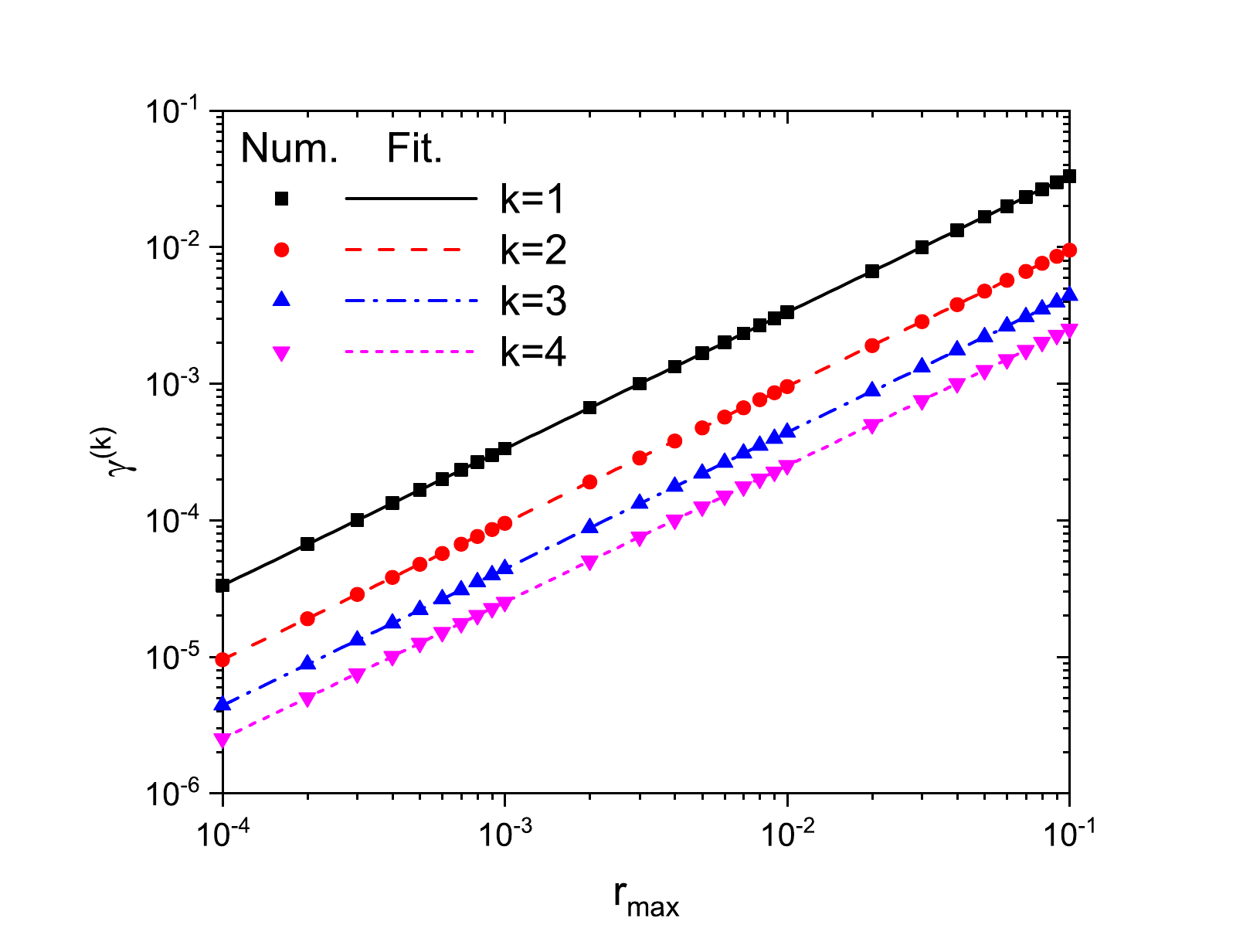}
\caption{\label{fig3} Asymptotic behavior of the $2^k$-pole nuclear shielding factors in the limit of $r_{\max} \to 0$ for the ground state of the spherically confined hydrogen atom. Dots and lines represent numerical calculations and linear fittings, respectively.}
\end{figure}

The asymptotic behavior of multipole nuclear shielding factors for the confined hydrogen atom at small confinement radii is another interesting topic that deserves in-depth studies. Figure \ref{fig3} displays the present numerical calculations of $\gamma^{(k)}$ for $r_{\max}$ ranging from $10^{-1}$ to $10^{-4}$. It is interestingly observed that all four nuclear shielding factors shown in the figure follow a simple linear law
\begin{equation}
\label{eq44}
\gamma^{(k)} (r_{\max}) \propto r_{\max},
\end{equation}
which is independent of the number of poles of the interaction. A simple explanation of such behavior can be built upon the fact that as $r_{\max}$ tends to zero the system wave functions become those of a particle in a spherical box with an ineffective constant potential. As a result, the ground state wave function approaches the limit of
\begin{equation}
\label{eq45}
\psi_{0}(r) = \frac{1}{\sqrt{2 \pi r_{\max}}} \frac{\sin (\pi r / r_{\max})}{r},
\end{equation}
and the corresponding radial expectation values $\left\langle r^p \right\rangle$ are given by
\begin{equation}
\label{eq46}
\left\langle r^p \right\rangle = r_{\max}^{p} \int_{0}^{1} 2 x^p \sin^2(\pi x) dx.
\end{equation}
It can be readily concluded that the radial expectation values are proportional only to the powers of confinement radius, i.e., $\left\langle r^p \right\rangle \propto r_{\max}^{p}$. The combination of Eq. (\ref{eq46}) with the first-order VPT approximation of the multipole nuclear shielding factor in Eq. (\ref{eq34}) or the second-order approximation in Eq. (\ref{eq35}) gives rise straightforwardly to a simple linear law with respect to $r_{\max}$.

\begin{figure*}[!htbp]
\includegraphics[width=0.365\textwidth]{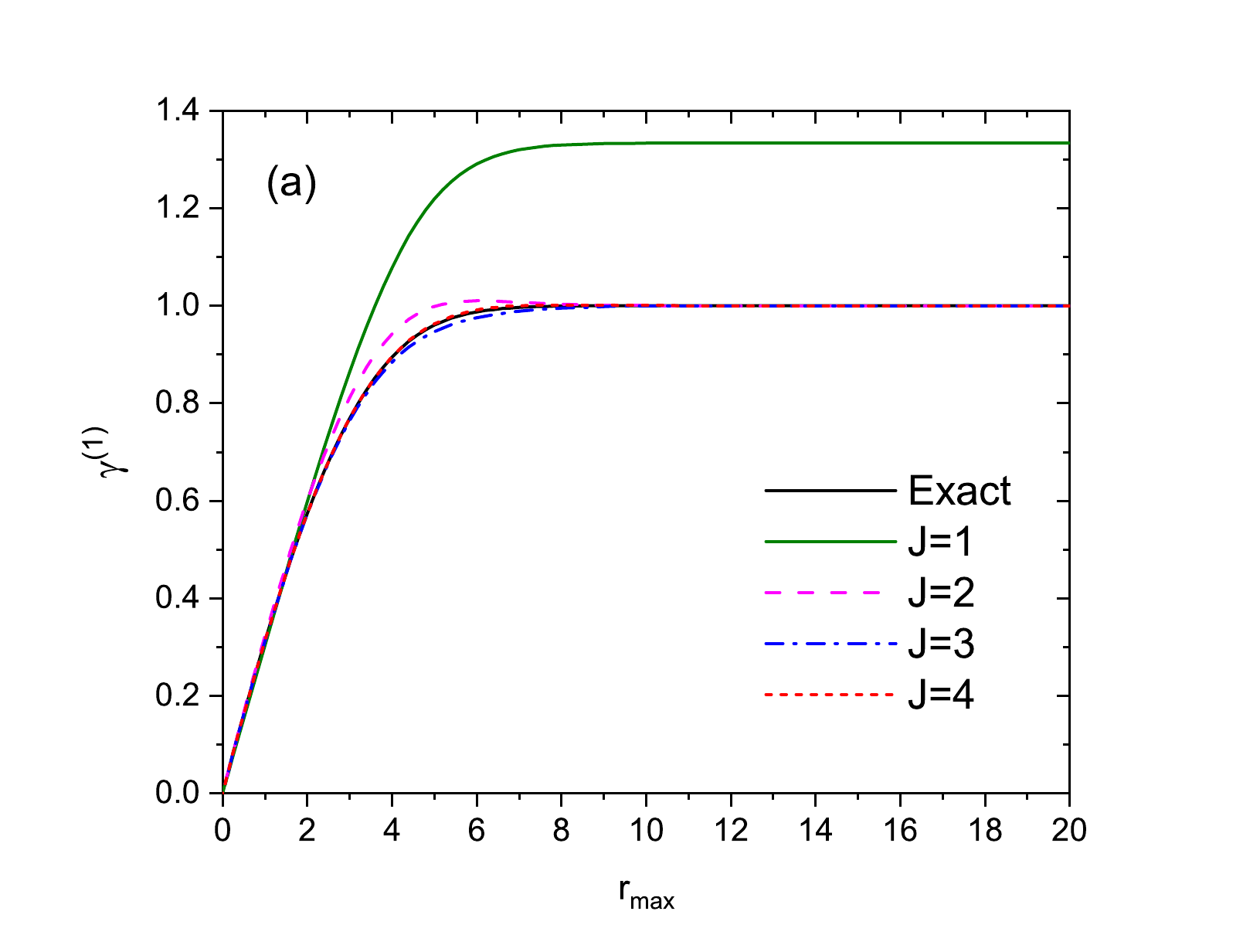} \includegraphics[width=0.365\textwidth]{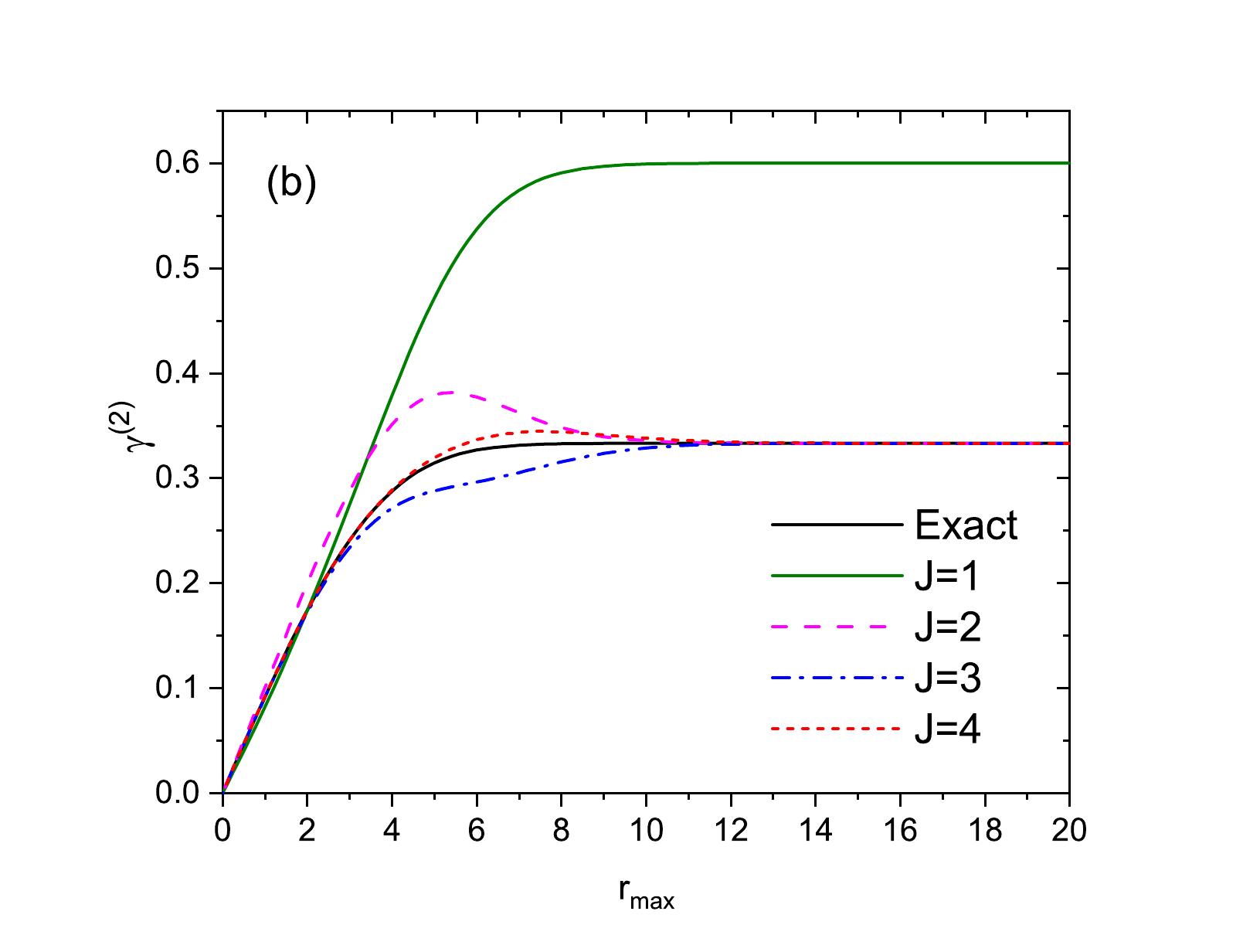} \\
\includegraphics[width=0.365\textwidth]{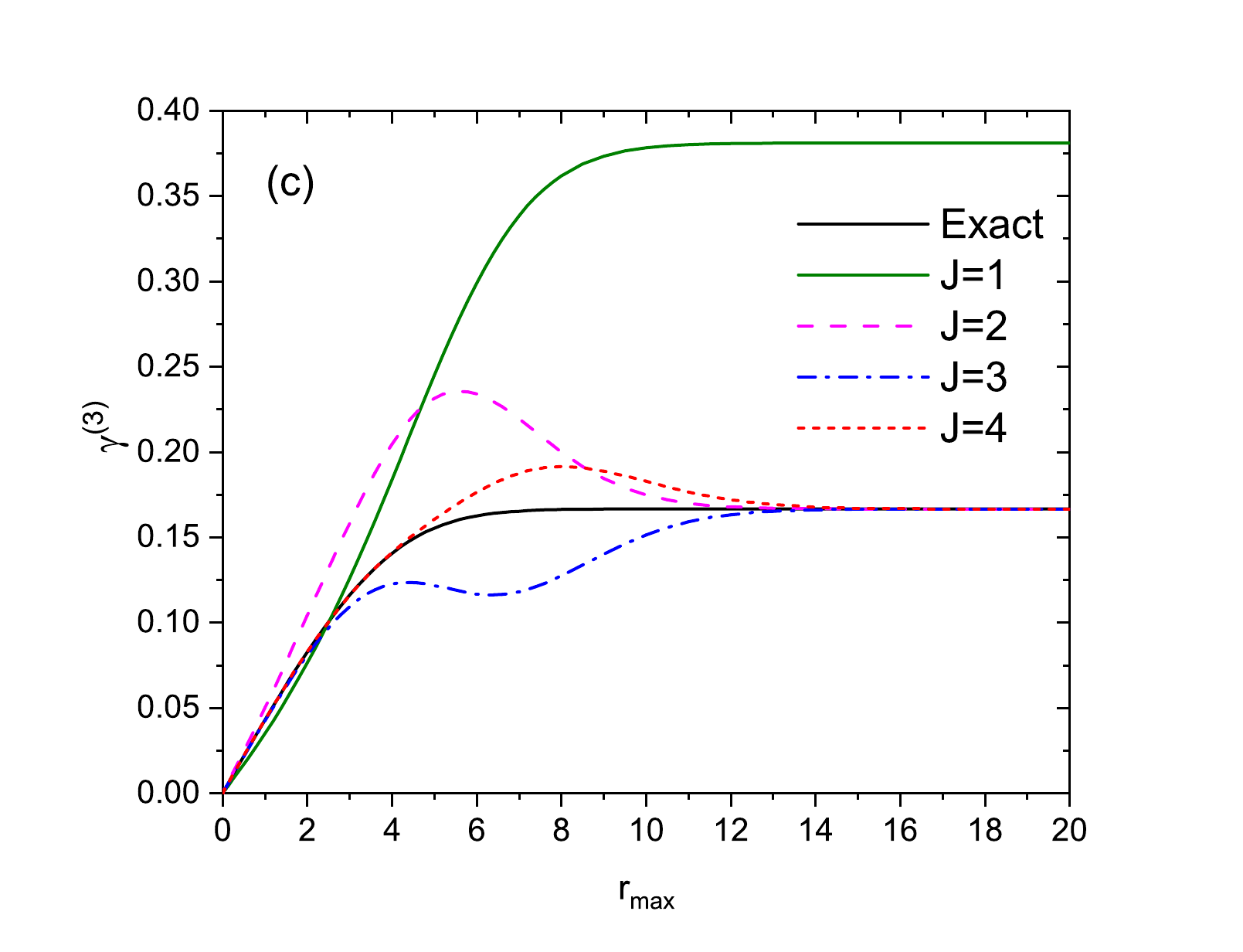} \includegraphics[width=0.365\textwidth]{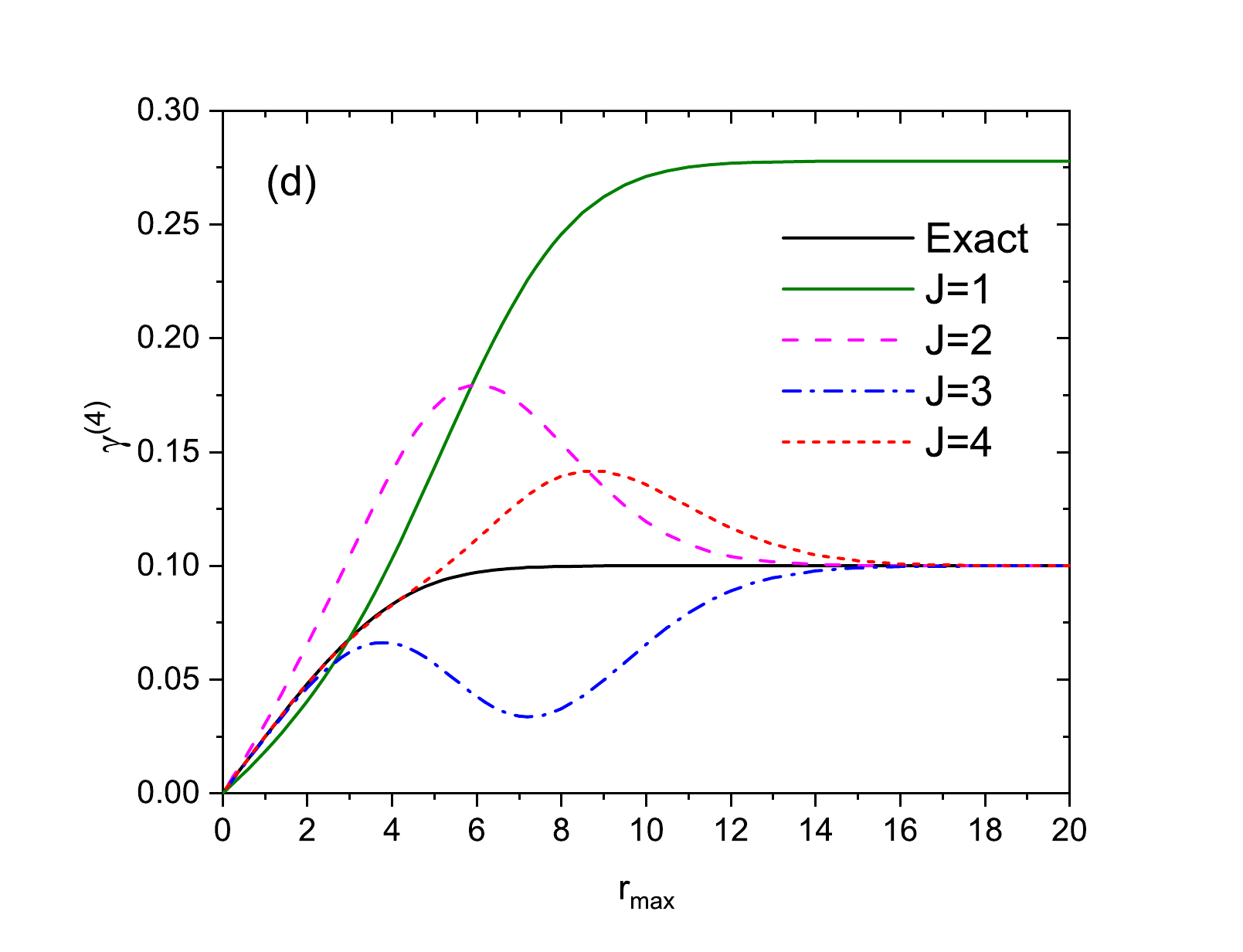}
\caption{\label{fig4} The $J$th-order ($J=1, 2, 3, 4$) VPT approximations of the $2^k$-pole ($k=1, 2, 3, 4$) nuclear shielding factors for the ground state of the confined hydrogen atom. Panels (a), (b), (c), and (d) represent the dipole, quadrupole, octupole and hexadecapole nuclear shielding factors, respectively. In each panel, the black solid line represents accurate sum-over-states calculations.}
\end{figure*}

\begin{figure*}[!htbp]
\includegraphics[width=0.365\textwidth]{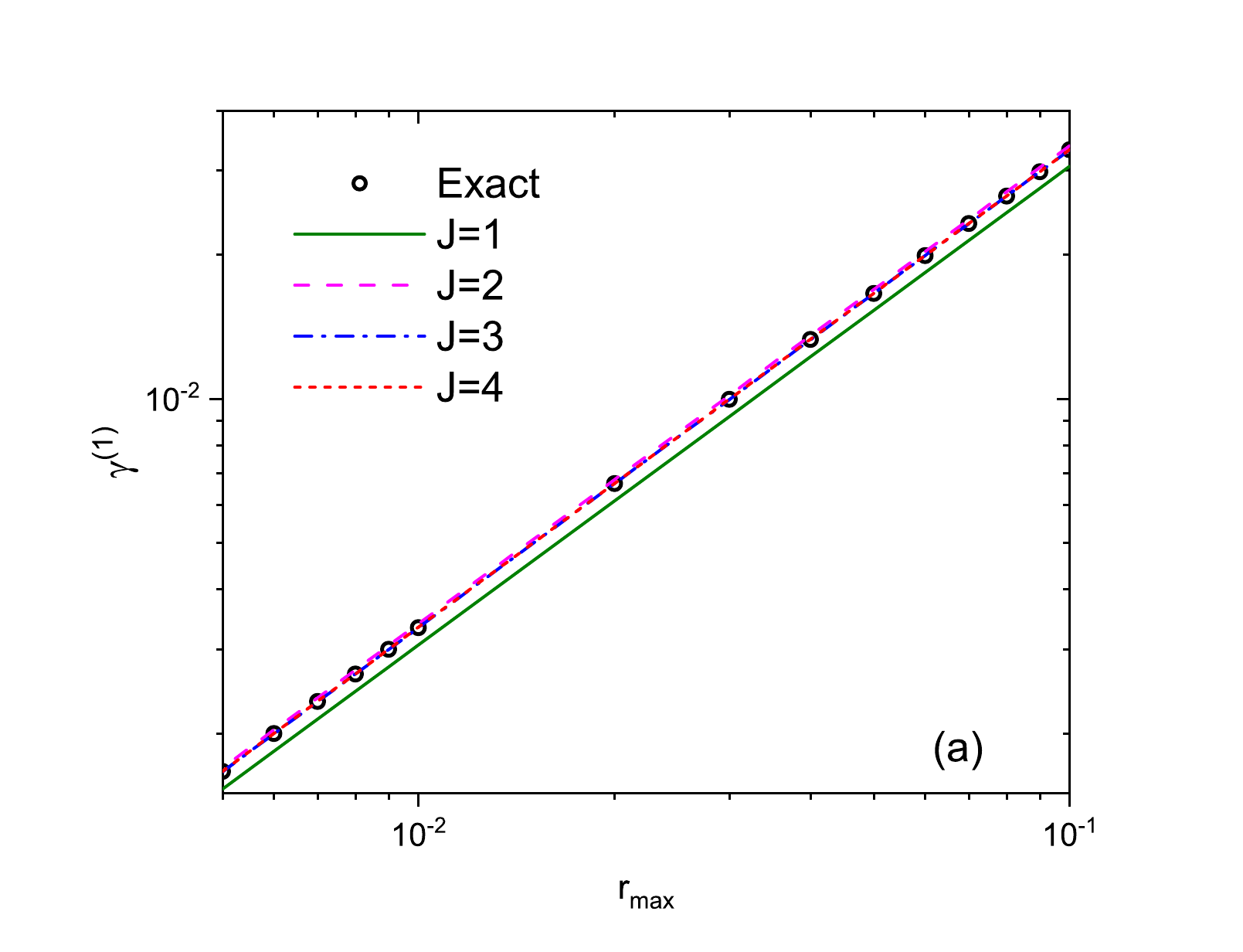} \includegraphics[width=0.365\textwidth]{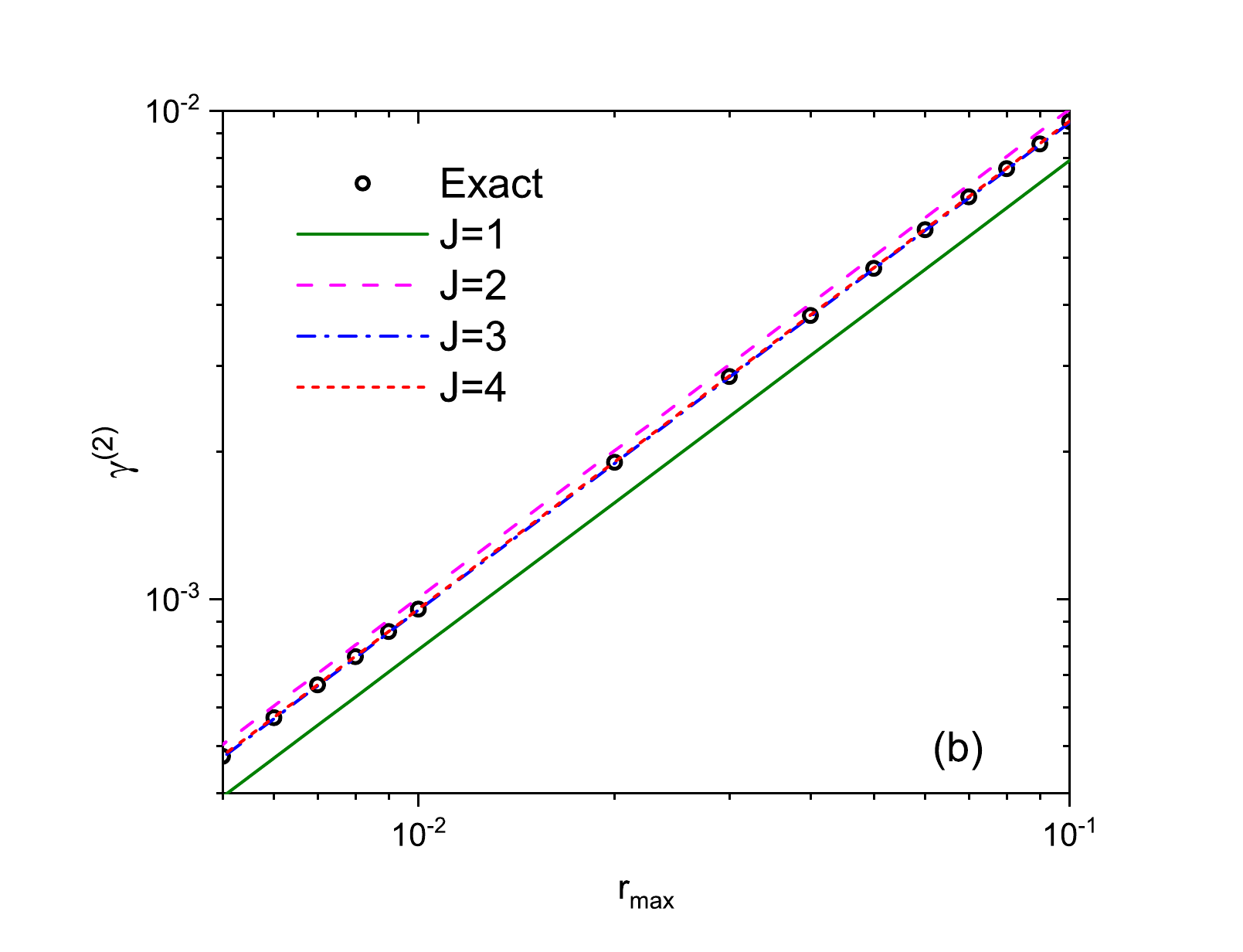} \\
\includegraphics[width=0.365\textwidth]{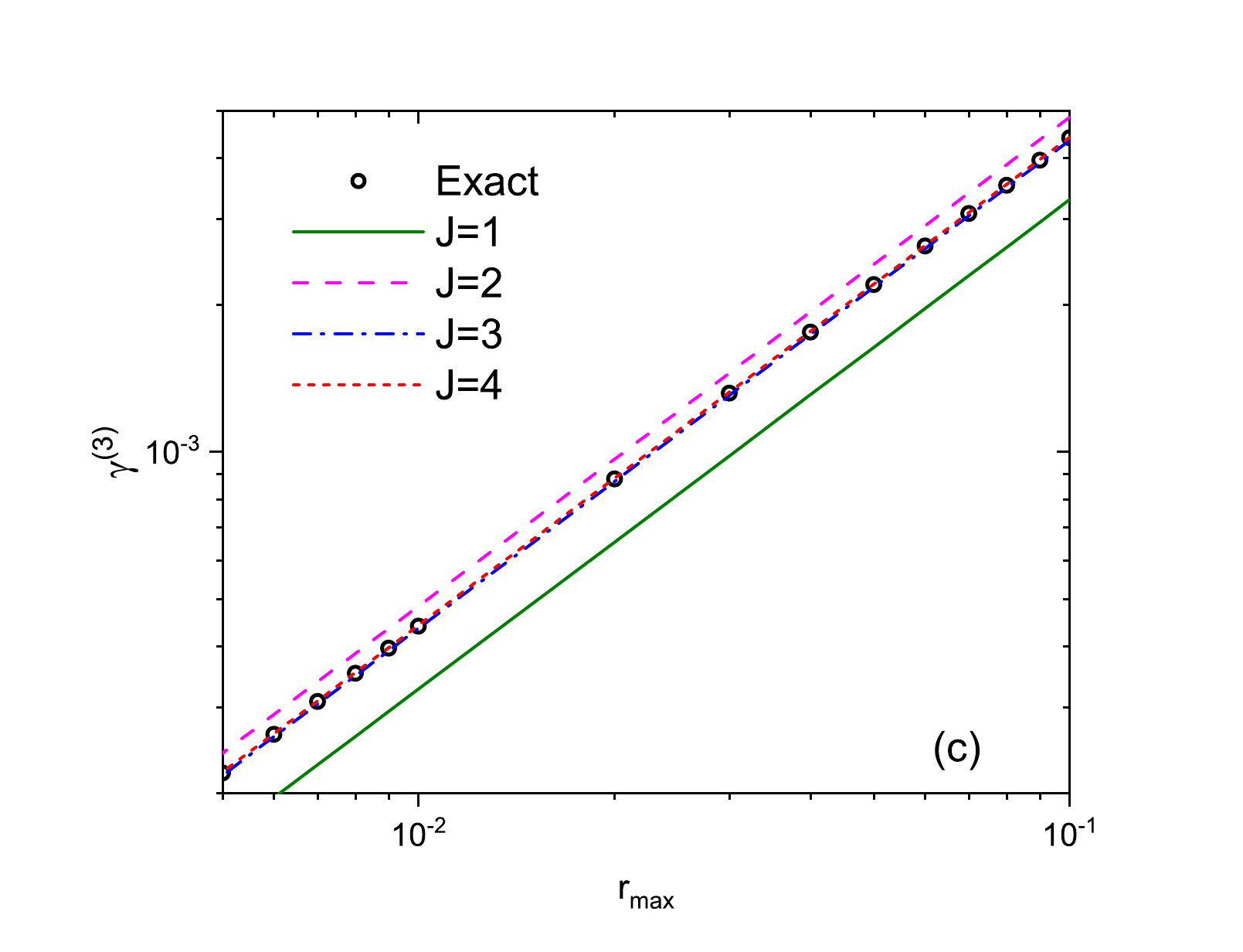} \includegraphics[width=0.365\textwidth]{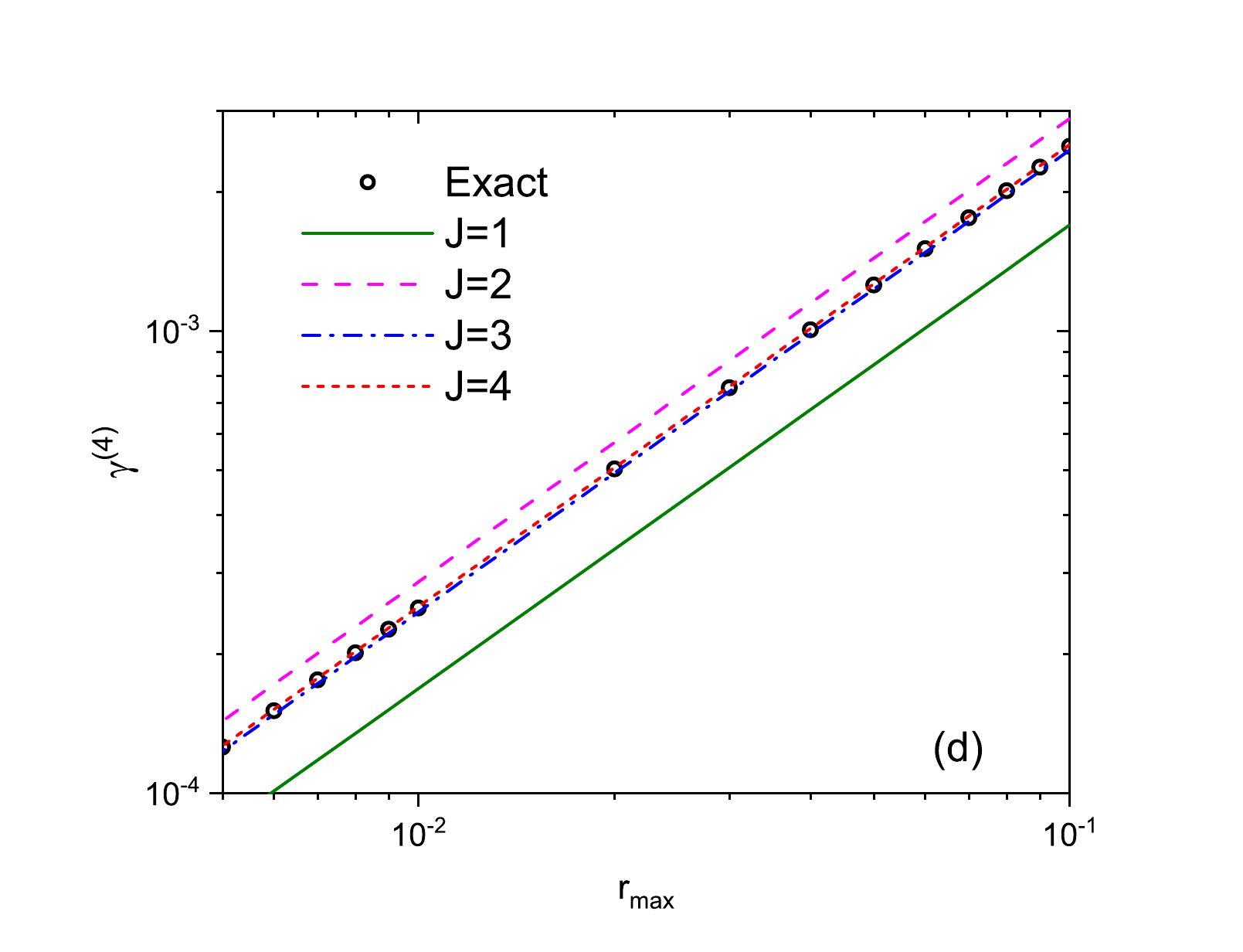}
\caption{\label{fig5} Same as Fig. \ref{fig4} but for the asymptotic behavior of the $2^k$-pole ($k=1, 2, 3, 4$) nuclear shielding factors at small values of $r_{\max}$. Panels (a), (b), (c), and (d) represent the dipole, quadrupole, octupole and hexadecapole nuclear shielding factors, respectively. In each panel, the black hollow dots represent accurate sum-over-states calculations.}
\end{figure*}

For the dipole interaction, Laughlin \cite{LaughlinAQC} has performed a small-box perturbation expansion of the ground state wave function and derived an expansion for the dipole nuclear shielding factor in the form
\begin{equation}
\label{eq47}
\gamma^{(1)} (r_{\max}) = 0.33333 r_{\max} - 0.01899 r_{\max}^2 + O \left( r_{\max}^3 \right).
\end{equation}
Such a high-order formula is indistinguishable from the dipole fitting curve drawn in Fig. \ref{fig3} due to the extremely small magnitude of the quadratic term. Actually, the combination of VPT approximations in Eqs. (\ref{eq36}) and (\ref{eq37}) with the radial expectation values in Eq. (\ref{eq46}) results in approximate linear laws \cite{FowlerMP}
\begin{equation}
\label{eq48}
\gamma_1^{(1)} (r_{\max}) \approx 0.30625 r_{\max}; \ \ \gamma_2^{(1)} (r_{\max}) \approx 0.33943 r_{\max}.
\end{equation}
The coefficient approaches the constant $1/3$ from the lower and upper sides alternatively with continuously increasing the order $J$.

The applicability of the Hylleraas VPT approximation in a wide range of confinement radius is examined in Fig. \ref{fig4}, where the $J$th-order approximations $\gamma_{J}^{(k)}$ ($J=1$ to $4$) are compared with the accurate numerical calculations collected in Table \ref{tab1}. All radial expectation values required in the VPT calculations are produced by the GPS method with the same parameters as mentioned above. It can be readily seen that the first-order approximation of $\gamma_{1}^{(k)}$ differs from the accurate results at large confinement radii, while higher-order approximations of $\gamma_{J}^{(k)}$ with $J \geq 2$ successfully rebuild the free-atom values as $r_{\max} \to \infty$. This is consistent with the discussion in Eqs. (\ref{eq39}) to (\ref{eq42}) that the second-order VPT approximation reproduces exactly the analytical solution of multipole nuclear shielding factors for the free hydrogen atom. Through an overall view of Fig. \ref{fig4} one can generally draw a conclusion that the applicability of the VPT method improves with increasing the order $J$, especially at small confinement radii. We show in Fig. \ref{fig5} the comparison of $\gamma_{J}^{(k)}$ ($J=1$ to $4$) with accurate numerical results for the confinement radius smaller than $0.1$. The agreement becomes better with increasing $J$ and the third-order approximation $\gamma_{3}^{(k)}$ is nearly indistinguishable from the numerical results at these small values of $r_{\max}$. Another distinct feature of Fig. \ref{fig5} is that the VPT approximation in arbitrary orders follows the same linear law of Eq. (\ref{eq44}) for all multipole nuclear shielding factors at sufficiently small confinement radii.

\begin{figure}[!tbp]
\includegraphics[width=0.5\textwidth]{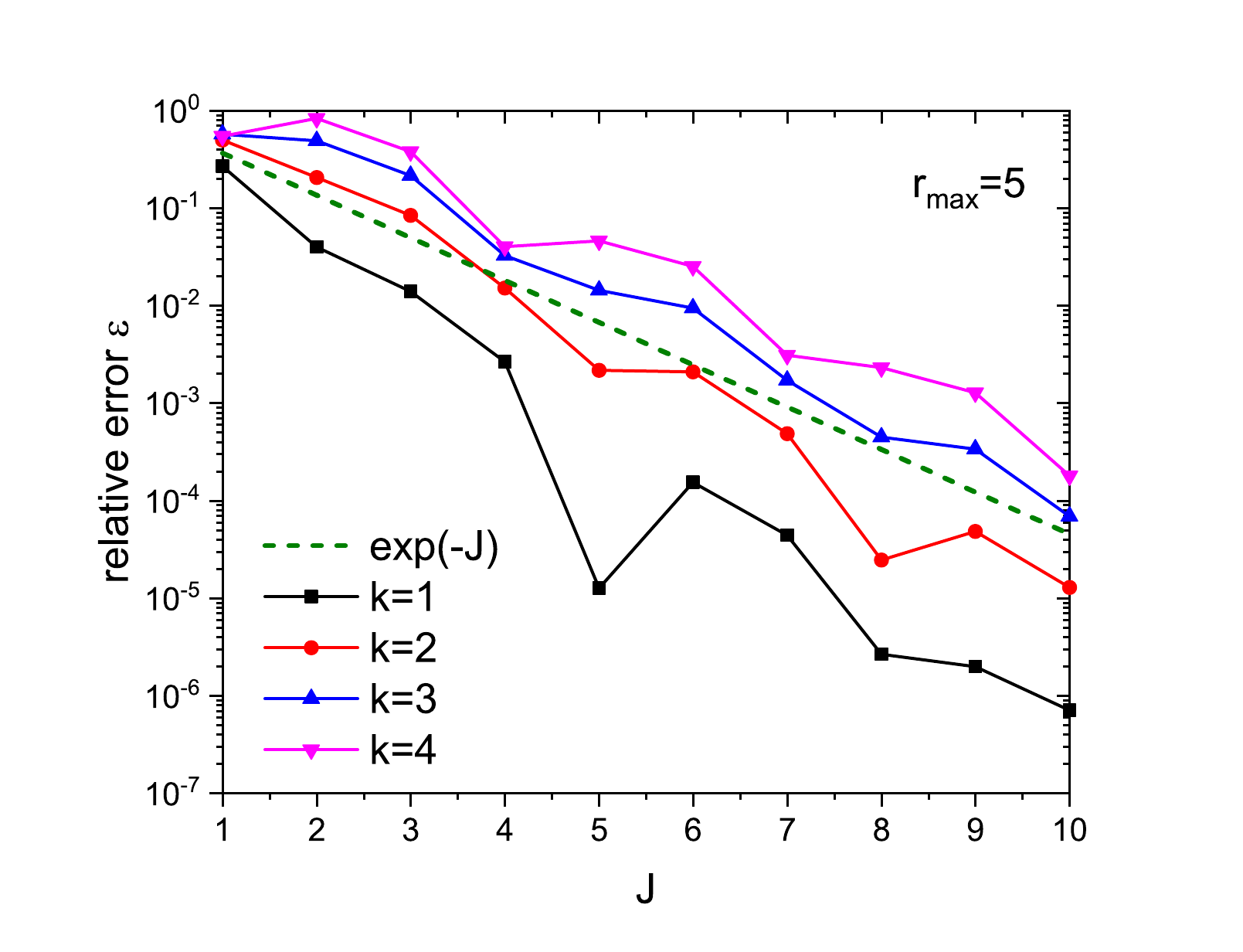}
\caption{\label{fig6} Relative errors of the VPT approximations of the $2^k$-pole ($k=1, 2, 3, 4$) nuclear shielding factors for the ground state of the confined hydrogen atom at $r_{\max} = 5$ with respect to increasing the order $J$. The dashed line indicates the empirical convergence rate of $\varepsilon = \exp(-J)$.}
\end{figure}

From Fig. \ref{fig4} one also notices that the VPT approximation becomes worse at intermediate confinement radii. This is probably because in such a situation, both the confinement potential and the Coulomb interaction are of crucial importance in modulating the system wave functions and, consequently, no simple approximation can be made on the wave functions as well as the nuclear shielding factors. Taking $r_{\max} = 5$ as an example, Fig. \ref{fig6} displays the relative errors of the VPT approximations defined by
\begin{equation}
\label{eq49}
\varepsilon (J) = \frac{\left| \gamma_{\text{Exact}}^{(k)} - \gamma_{J}^{(k)} \right|}{\gamma_{\text{Exact}}^{(k)}},
\end{equation}
for the multipole nuclear shielding factors with $J$ increasing from $1$ to $10$. Also included in this figure is an empirical exponential convergence rate for the VPT approximation \cite{XiaPRE} with increasing $J$
\begin{equation}
\label{eq50}
\varepsilon (J) \propto e^{-J}.
\end{equation}
From Eq. (\ref{eq28}) we know that the increase of the order of the VPT approximation provides a more complete basis set to construct the perturbed wave function. Therefore, the order $J$ plays the same role as the dimension of the basis set in describing the corresponding subspace, leading to an approximate exponential convergence as in the basis-expansion methods for one-electron atoms \cite{YuFBS}. From Fig. \ref{fig6} one finds that the convergence rates for different poles of nuclear shielding factors are slightly different, with the higher-pole term converging slower. This is probably because the $2^k$-pole perturbation potential $r^k P_k(\cos \theta)$ projects the first-order correction of the $s$-wave ground state wave function of the confined hydrogen atom into a subspace expanded by $k$-wave functions. As a result, a larger basis set is required to construct the perturbed wave function with higher angular momentum localized in a large region.

\section{Conclusion} \label{sec4}

In this work, we investigated the multipole nuclear shielding factors of one-electron atoms by developing the sum-over-states numerical method and the Hylleraas VPT approximations. One of the great advantages of these two approaches rests upon the fact that there is no need to calculate the perturbed wave functions of the atom under the interaction of an external field. The sum-over-states method only needs the transition matrix elements between the unperturbed eigenstates and the Hylleraas VPT approximation just requires the radial expectation values of the initial ground state. All these quantities for the one-electron atoms can be efficiently and accurately calculated via the sophisticated GPS numerical method.

The two methods developed were employed to calculate the multipole nuclear shielding factors for the ground state of the spherically confined hydrogen atom. The present dipole results are in excellent agreement with previous predictions, while the higher-pole results are reported for the first time. It was interestingly found that all multipole nuclear shielding factors tend to zero following a linear law as the confinement radius approaches zero. Such behavior has been explained by utilizing the VPT approximations together with the model of a particle in a spherical box. The Hylleraas VPT method shows comparably slow convergence at intermediate confinement radii where both the confinement potential and the Coulomb interaction are of crucial importance in modulating the system wave functions. The convergence analysis in this region reveals that the VPT method manifests an approximate exponential convergence rate with increasing orders. It is of great interest in the future to investigate the multipole nuclear shielding factors and corresponding asymptotic laws for multi-electron confined systems.

\section*{Declaration of Interest Statement}

The authors declare that they have no known competing financial interests or personal relationships that could have appeared to influence the work reported in this paper.

\section*{Data Availability}

The data that support the findings of this study are available within the article.

\newpage

\section*{Acknowledgements}

This work was supported by the National Key Research and Development Program of China (Grant No. 2022YFE0134200) and the National Natural Science Foundation of China (Grant No. 12174147).


\nocite{*}
\bibliography{nsf-ch}

\end{document}